\documentclass[twocolumn, showpacs, longbibliography, prx,aps, superscriptaddress, amsmath, amssymb]{revtex4-1}
\usepackage{graphicx}
\usepackage{dcolumn}
\usepackage{bm}

\usepackage[T1]{fontenc}
\usepackage[colorlinks,linkcolor=blue,citecolor=blue,urlcolor=blue]{hyperref}

\begin{document}

\title{From weak to strong coupling of localized surface plasmons to guided modes in a luminescent slab}

\author{S.R.K. Rodriguez}
\affiliation{Center for Nanophotonics, FOM Institute AMOLF, c/o
Philips Research Laboratories, High Tech Campus 4, 5656 AE
Eindhoven, The Netherlands} \email{s.rodriguez@amolf.nl}

\author{Y.T. Chen}
\affiliation{School of Optical and Electronic Information, Huazhong University of Science and Technology, Wuhan, 430074, China}

\author{T.P. Steinbusch}
\affiliation{Center for Nanophotonics, FOM Institute AMOLF, c/o Philips Research Laboratories, High Tech Campus 4, 5656 AE Eindhoven, The Netherlands}

\author{M.A. Verschuuren}
\affiliation{ Philips Research Laboratories, High Tech Campus 4, 5656 AE Eindhoven, The Netherlands}

\author{A. F. Koenderink}
\affiliation{Center for Nanophotonics, FOM Institute AMOLF, Science Park 104, 1098 XG Amsterdam, The Netherlands}

\author{J. G\'{o}mez Rivas}
\affiliation{Center for Nanophotonics, FOM Institute AMOLF, c/o Philips Research Laboratories, High Tech Campus 4, 5656 AE Eindhoven, The Netherlands}
\affiliation{COBRA Research Institute, Eindhoven University of Technology, P.O. Box 513, 5600 MB Eindhoven, The Netherlands}

\begin{abstract}
We investigate a periodic array of aluminum nanoantennas embedded in a light-emitting slab waveguide.
By varying the waveguide thickness we demonstrate the transition from weak to strong coupling between localized surface plasmons in the nanoantennas and refractive index guided modes in the waveguide. We experimentally observe a non-trivial relationship between extinction and emission dispersion diagrams across the weak to strong coupling transition.  These results have implications for a broad class of photonic structures where sources are embedded within coupled resonators. For nanoantenna arrays, strong vs. weak coupling leads to drastic modifications of radiation patterns without modifying the nanoantennas themselves, thereby representing an unprecedented design strategy for nanoscale light sources.
\end{abstract} \pacs{73.20.Mf, 42.82.Et, 71.36.+c, 33.50.Dq}

\narrowtext \maketitle

Coupled systems are ubiquitous in physics. In recent years, the design and description of coupled nanoscale optical resonators has been greatly inspired by the field of atomic physics. Strong and weak coupling phenomena have been reported for light-driven molecular, metallic, and dielectric nanoscale systems. In the weak coupling regime,  lineshapes akin to Fano resonances~\cite{Fano} and  Electromagnetically Induced Transparency (EIT)~\cite{Harris91} have attracted much attention~\cite{Sarrazin03, Genet03, Zhang08, Giessen09, Soukoulis09, Halas10, Kekatpure&Brongersma10, FanoRev10, FanoNat, Sassan}. Both of these effects arise from the interference between spectrally broad and narrow resonances, while the energy detuning sets them apart (zero-detuning for EIT vs. large detuning for Fano resonance). Interference can lead to a pronounced spatial and angular redistribution of optical states~\cite{Alonso11, Frimmer12}, which has important implications for sensing~\cite{Nordlander08, Neubrech08, Offermans}, and enhanced spontaneous emission~\cite{Vecchi09, Rodriguez12APL, Rodriguez12PRL, Lozano13}. On the other hand, the strong coupling regime --- wherein the energy exchange rate between the coupled modes exceeds their loss rates --- has been observed in various systems combining photons, excitons, and/or surface plasmon polaritons~\cite{Bellessa2004, Dintinger2005, Vasa2008, Manjavacas, Schwartz11, Gonzalez-Tudela2013, Rodriguez13PRL, Rodriguez13OE, Torma14}. Advantageously, strong coupling enables to significantly modify the optical and chemical properties of the participating systems~\cite{Hutchison12, Hutchison13}. This follows from the fact that the properties of strongly coupled states are intermediate to those of the bare states.

In this manuscript, we demonstrate how localized surface plasmons in the same nanoantenna array transition from weak to strong coupling with a refractive index guided mode in  a luminescent slab. Nanoantennas provide an interface between plane waves in the far-field and localized energy in the near-field~\cite{Palash09}, while dielectric waveguides can guide this energy to a desired position with low losses. Therefore,  understanding the conditions enabling an efficient coupling between these two photonic building blocks is an important endeavour in optics. Indeed, several theoretical and experimental works have demonstrated that light can be received, transferred, or emitted, in unconventional ways when metallic resonators are either strongly or weakly coupled to dielectric waveguides ~\cite{Giessen03, Zentgraf09, Yannopapas09, Fevrier12, Rodriguez12PRL, Bernal12}. However, the transition from weak to strong coupling between the same two nanoantenna and waveguide modes remains unexplored. Here we demonstrate this transition by varying the thickness of a polymer waveguide within which a metallic nanoantenna array is embedded. We demonstrate the impact of this transition on the variable angle light extinction and emission spectra of the system. The emission stems from luminescent molecules embedded in the waveguide. We find that an optimum waveguide thickness exists for increasing the ratio of the coupling rate to the loss rates, thereby providing a design principle for accessing the strong coupling regime.  Finally, we discuss  differences between the light emission and extinction spectra across the weak-to-strong coupling transition, and we explain their origin on the transmutation of coupled optical modes with varying degree of field confinement.

\begin{figure}[tb]
\includegraphics[width=\linewidth]{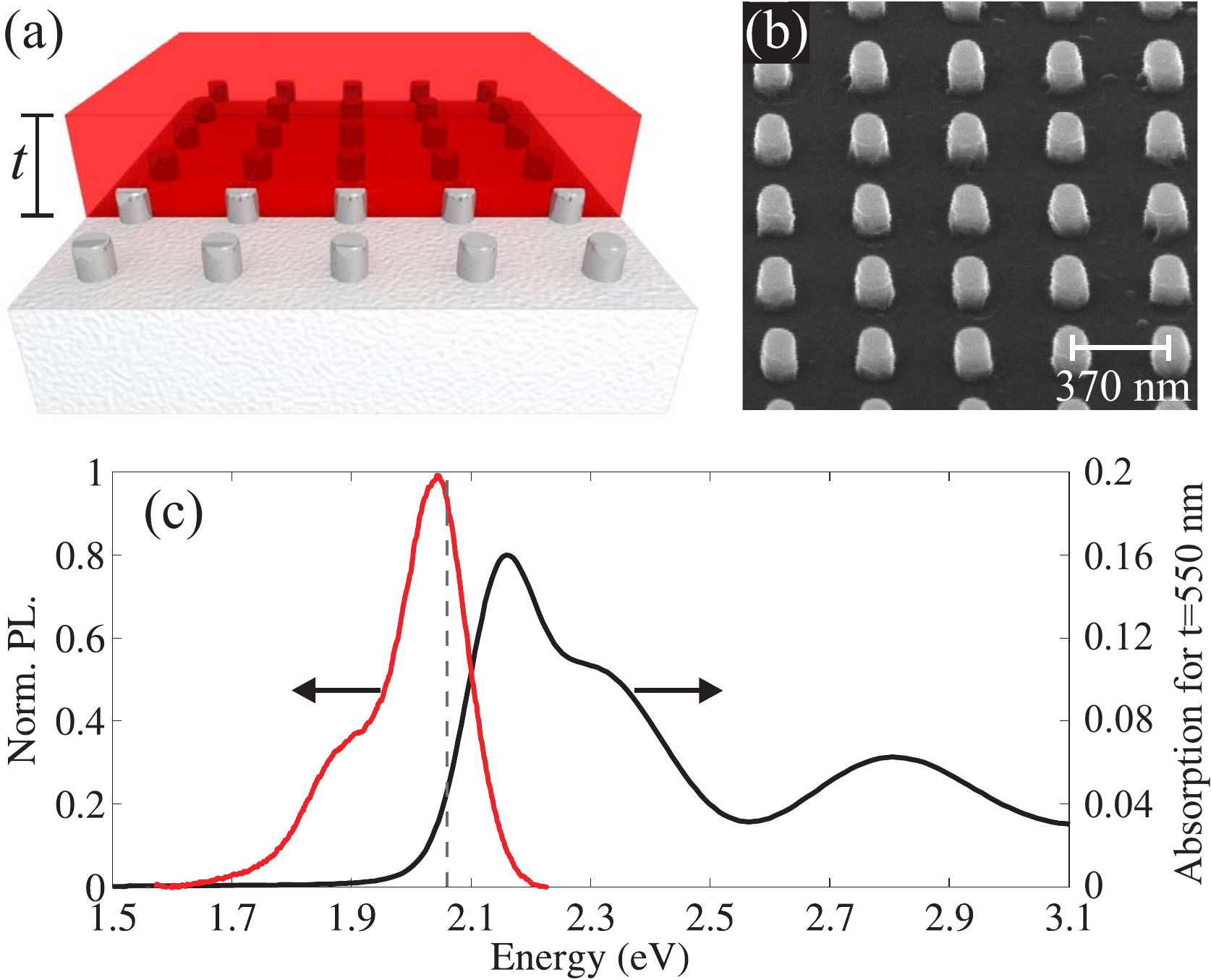}
\caption{(a) 3D schematic representation of the sample. An aluminum nanoantenna
array stands on a SiO$_2$ substrate, and is embedded in a
luminescent slab waveguide of thickness $t$. (b) Inclined-view ($43^\circ$ off the normal) scanning electron micrograph of the nanoantenna array prior to the deposition of the waveguide. (c) Absorptance spectrum of a $t=550$ nm dye-doped waveguide, and normalized photoluminescence spectrum.  The dashed line indicates the energy at which the fundamental TM$_0$ guided mode and the bare localized surface plasmon resonance cross. }\label{fig1}
\end{figure}

Figure~\ref{fig1}(a) illustrates the sample. An aluminum nanoantenna array with a total size of $2 \times 2$ mm$^2$ was fabricated by substrate conformal imprint lithography~\cite{SCIL} and reactive ion etching of aluminium onto a fused silica substrate. Figure~\ref{fig1}(b) shows an inclined view ($43^\circ$ off the normal) scanning electron micrograph of the array. The nanoantennas are approximately disks with a diameter of $130 \pm 20$ nm and a height of $150 \pm 10$ nm, arranged in a square lattice with a constant $a = 370 \pm 5$ nm. On top of the array we spin-coated a toluene solution with polystyrene and the organic dye Lumogen F305. Consequently, the toluene evaporated leaving a dye-doped polystyrene layer. The refractive index of this layer is higher than the underlying silica and overlying air, such that the array is embedded in a slab waveguide. We varied the thickness $t$ of this waveguide by controlling the spin-rate during the deposition and the viscosity of the solution. The latter was controlled through the polystyrene-to-toluene ratio, while the dye-to-polystyrene ratio (determining the final molecular concentration in the waveguide) was held constant at 3 weight $\%$ .   The influence of this molecular concentration on the optical properties of the waveguide  was assessed  through ellipsometry measurements, which yielded the complex refractive index $\tilde{n}_p = n_p + i k_p$ of the dye-doped polymer. Over the entire visible spectrum, $n_p$ varied less than $0.8 \%$ with respect to the index of a polystyrene layer without molecules. $k_p$ determines the absorptance of the waveguide, which we plot in Fig.~\ref{fig1}(c) for $t=550$ nm as a black line. The absorptance is defined as $1-I/I_0$ where $I/I_0 =$ exp$\left( -4\pi k_p t/ \lambda \right)$ with $I_0$ the incident intensity, $I$ the intensity transmitted through the dye layer, and $\lambda$ the free-space wavelength. At the energy of the dashed line in Fig.~\ref{fig1}(c) (where the metallic nanoantennas and waveguide are tuned in resonance, as explained ahead), only $4.6 \%$ of the incident light is absorbed by the molecules. This allows us to exclude the influence of the molecules on the nanoantenna-waveguide coupling at this energy. The  photoluminescence spectrum of the waveguide is shown as a red line in Fig.~\ref{fig1}(c).


\begin{figure*}[!]
\includegraphics[width=\linewidth]{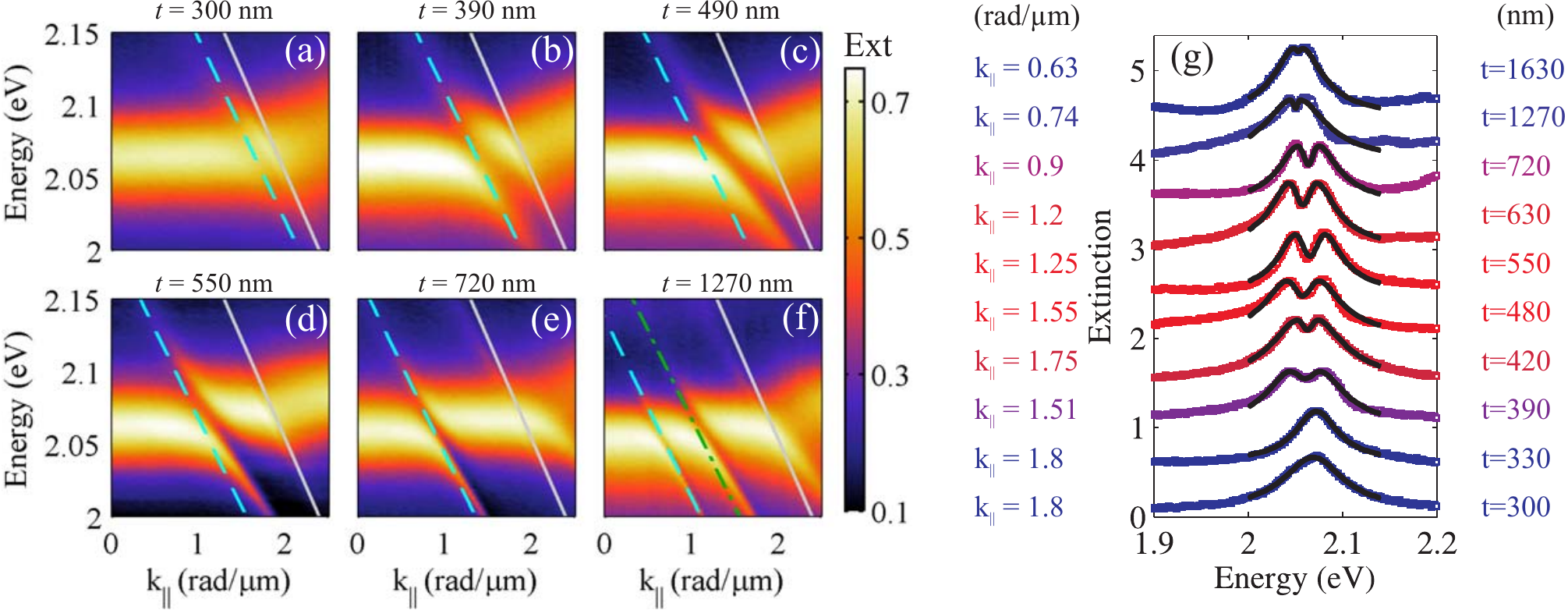}
\caption{ Extinction measurements of the structure in Fig.~\ref{fig1}, for a waveguide thickness (a) $t=300$ nm, (b) $t=390$ nm,  (c) $t=490$ nm, (d) $t=550$ nm, (e) $t=720$ nm, and (f) $t=1270$ nm.  The gray solid line, identical for all plots, indicates the Rayleigh anomaly in the substrate. The cyan dashed line, shifting towards lower $k_\|$ for increasing $t$, indicates the fundamental TM$_0$ guided mode. The green dash-dotted line in (f) indicates TM$_1$ guided mode. (g) cuts in $k_\|$ at zero detuning between the TM$_0$ guided mode and the localized surface plasmon resonance in arrays embedded in waveguides of different thickness. For each cut, the value of  $k_\|$ and $t$ is indicated at the left and right of the figure, respectively, in the same color as the measurement. The black lines overplotted with the measurements are fits with a coupled oscillator model as described in the text. For successive increments in $t$, the extinction is increased by 0.5 for clarity. }\label{fig2}
\end{figure*}

Figure~\ref{fig2} shows a series of extinction measurements of the same nanoantenna array embedded in waveguides of different thickness. The sample is illuminated by a collimated (angular spread $< 0.1^{\circ}$) TM-polarized white light beam from a halogen lamp, while a fiber-coupled spectrometer collects the transmitted light in the far-field.  The extinction is defined as 1-$T_0$, with $T_0$ the zeroth-order transmission through the array normalized to the transmission through the dye-doped polymer layer and substrate. We plot the extinction in color --- same scale for all plots --- as a function of the incident photon energy and in-plane momentum $k_\|$. A computer-controlled stage was used to rotate the sample by an angle $\theta_{in}$,  thereby changing the component of the wave vector parallel to the lattice vector, i.e., $\mathbf k_\|= k_0 \sin(\theta_{in}) \bm a$. $k_0$ is the magnitude of the free space wave vector and  $\bm a$  is a unit vector parallel to one of the two equivalent lattice vectors. We refer to the magnitude of $\mathbf k_\|$ as $k_\|$. The angular resolution of the measurements is $0.2^{\circ}$. We focus on TM polarization because excellent spectral overlap between the coupled modes and the emission from the dye molecules aids to bring out the hybridization effects in both emission and extinction of TM-polarized light. However,  strong coupling is not particular for one polarization, as confirmed for example by  experiments with TE-polarized light~\cite{Rodriguez12PRL}.

We now interpret the various features observed in the measurements in Fig.~\ref{fig2}. For all $t$, the broad extinction peak near 2.07 eV  with a flat angular dispersion at small $k_\|$ corresponds to the excitation of localized surface plasmon resonances (LSPRs). A plane wave that excites LSPRs can also be diffracted grazing to the surface of the array, leading to the so-called Rayleigh anomaly (RA) condition. The gray solid line overplotted on the measurements in Fig.~\ref{fig2} indicates the RA in glass, with a  dispersion given by $E_{R\pm}(k_\|) = \frac{\hbar c}{n_g}  |  k_\|  + m G |$. Here, $m=-1$ is the relevant order of diffraction, $G = \frac{2 \pi}{a}$ is the magnitude of the reciprocal lattice vector,  and $n_{g}=1.44$ is the refractive index of the glass substrate. The periodic array may also enable the plane wave excitation of a guided mode in the polymer layer, which has a refractive index higher than its surroundings.  The cyan dashed line, changing with $t$, indicates the dispersion relation of fundamental TM$_0$  guided mode calculated using  the formalism described by Yariv and Yeh~\cite{Yariv}. We solve for the bound modes in a dielectric slab with refractive index $n_{p} = 1.58$ (polystyrene), sandwiched between semi-infinite media with $n_{a}=1.0$ (air) and  $n_{g}=1.44$ (glass). The thickness $t$ of the slab is obtained from profilometry measurements of the dye-doped polystyrene layer in experiments.

Figure~\ref{fig2} shows several dispersive features in extinction crossing or anti-crossing with the LSPR depending on $t$. The feature near the RA condition remains as a small perturbation on the LSPR for all $t$, and we therefore not dwell on it further.  We focus on the feature near the LSPR-TM$_0$-guided-mode crossing, which varies pronouncedly with $t$.  For $t=300$ nm, Fig.~\ref{fig2}(a) shows  a  weak narrow feature crossing with the LSPR without significantly affecting it.  This thin waveguide is close to cut-off, so the weakly confined TM$_0$ guided mode dispersion follows closely the RA dispersion.  As $t$ increases, the guided mode shifts away from the RA towards lower $k_\|$, and its signature in the spectra is clearly distinguished from the RA feature. In Fig.~\ref{fig2}(b) we begin to see signatures of hybridization between the LSPR and TM$_0$ guided mode. For increased $t$ [Figs.~\ref{fig2}(c,d)], a mode splitting emerges near zero detuning, where the energies of the bare LSPR and TM$_0$ guided mode cross but the coupled modes anti-cross. As $k_\|$ transits across the zero detuning point, the coupled modes gradually exchange their resemblance to one or the other of the bare modes. This adiabatic mode exchange across the zero detuning point is, qualitatively speaking, the signature of strong coupling. The hybrid excitations emerging from the strong LSPR-guided mode coupling are known as waveguide-plasmon polaritons~\cite{Giessen03, Rodriguez12PRL}. For $t=720$ nm [Fig.~\ref{fig2}(e)], the energy splitting between the same two modes is reduced, and for $t=1270$ nm [Fig.~\ref{fig2}(f)] the splitting is much smaller than the linewidths (weak coupling). For $t=1270$, the higher order TM$_1$ guided mode [green dash-dotted line in Fig.~\ref{fig2}(f)] is also excited. However, we do not observe indications of strong coupling between the TM$_1$ guided mode and the LSPR for any $t$.

An interesting observation in the dispersion diagrams in  Fig.~\ref{fig2} is that the calculated TM$_0$ guided mode and the corresponding feature in extinction are in better agreement for  thicker [Figs.~\ref{fig2}(e,f)] than for the thinner [Figs.~\ref{fig2}(a,b,c,d)] waveguides. We believe that this is due to the perturbation  of the ``bare'' waveguide structure by the nanoantennas. For thinner waveguides a higher fraction of the dielectric slab is occupied by the nanoantennas. Therefore, the actual structure deviates more pronouncedly from the planar layer considered in the calculations. The most significant deviations between the calculated TM$_0$ guided mode and the corresponding feature in extinction are observed  for the structures displaying the strongest splittings [Figs.~\ref{fig2}(c,d)], likely because in these cases the perturbative particle  has a greater overlap with the guided mode eigenfield.

Next, we analyze in Fig.~\ref{fig2}(g) the extinction measurements for various $t$ [more values than shown in  Fig.~\ref{fig2}(a)-(f)] at the value of $k_\|$ corresponding to zero LSPR and TM$_0$ guided mode detuning. This value of $k_\|$ (shown on the left of each plot) was established on the basis of a non-linear least squares fit of a model system --- coupled harmonic oscillators --- to the data, as we explain next.  In matrix form, the equations of motion of the model system are,
\begin{widetext}
\begin{equation}\label{COM}
\left( \begin{matrix}
\omega_{L}^2 - \omega^2  - i  \gamma_L \omega     & \Omega \omega  \\
\Omega \omega &  \omega_{G}^2(\bm k_\|) - \omega^2  - i  \gamma_G \omega 
\end{matrix} \right)
\left( \begin{matrix} x_L \\ x_G \end{matrix} \right) =
\left( \begin{matrix}\frac{F}{m} e^{-i\omega t} \\ 0  \end{matrix} \right),
\end{equation}
\end{widetext}
\noindent where we have assumed time-harmonic solutions. $\omega_{L}$ and $\omega_{G}(\bm k_\|)$ are the eigenfrequencies of the LSPR and TM$_0$ guided mode, $\gamma_L$ and $\gamma_G$ are their respective loss rates, while $x_L$ and $x_G$ are the oscillator displacements from equilibrium. $\Omega \omega$ represents the coupling strength between the two oscillators.  On the right hand side of Eq.~\ref{COM} appears the driving force per unit mass, $\frac{F}{m} e^{-i\omega t}$, which represents the incident optical field with frequency $\omega$. This force drives directly the LSPR only  because in the absence of scatterers, the guided mode is not directly driven by a plane wave incident from the far-field. The guided mode is  excited indirectly through the array. Our model assumes frequency-independent dissipative and coupling terms, which is valid for restricted energy ranges only. While relaxing these constraints could lead to a better quantitative agreement with the experiments, we show that a good fit and a reasonable description emerge in the spectral region of interest despite these simplifications.

\begin{figure}[!]
\includegraphics[width=\linewidth]{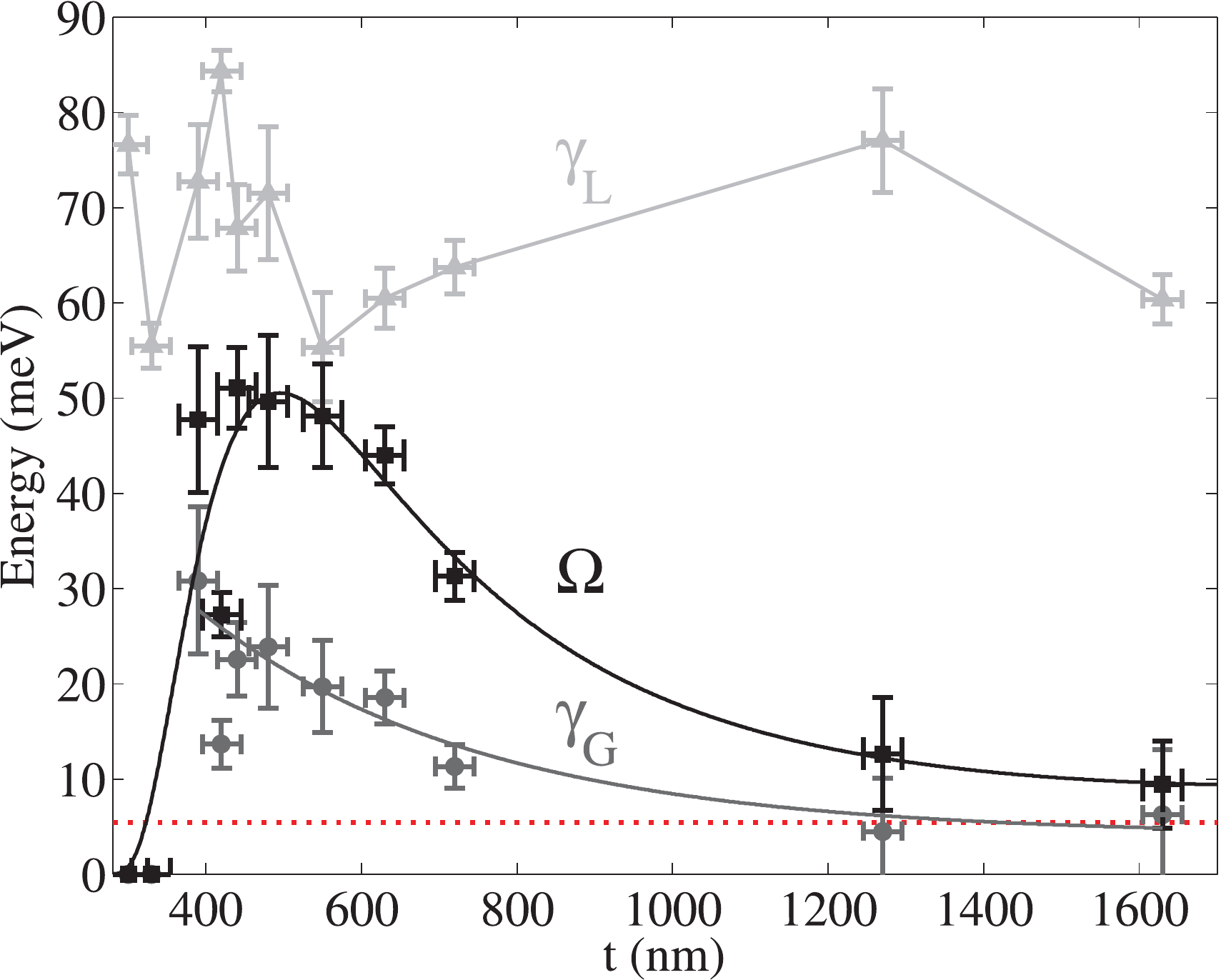}
\caption{ Coupling and loss rates extracted from the coupled oscillator model (Equation~\ref{COM}) fits to the extinction measurements as shown in Fig.~\ref{fig2}(g). Black squares are coupling rates, dark gray circles are loss rates of the TM$_0$ guided mode, and light gray triangles are loss rates for the localized surface plasmon resonance. Error bars in energy represent a 2$\sigma$ confidence interval on the fits. Error bars in thickness represent the uncertainty in the measurements of the waveguide thickness. The continuous lines overplotted with the data points are guides to the eye. The horizontal red dotted line indicates the absorption rate of the molecules in the waveguide at the average zero detuning energy.}\label{fig3}
\end{figure}

To establish the zero-detuning values of $k_\|$  we first let $\omega_G$ and $\omega_L$ be independent fit parameters. We fit the total power dissipated by both oscillators to the extinction spectra at the various values of $k_\|$. Zero-detuning is identified as the value of $k_\|$ for which the difference between $\omega_G$ and $\omega_L$ is minimized. Having established this value,  we then fit the model to the selected value of $k_\|$ once more, but now with the strict equality $\omega_G = \omega_L$. The black lines in Fig.~\ref{fig2}(g) are these fits. The model spectra capture the behavior in our measurements well. From the fits, we retrieve the coupling and loss rates as a function of $t$, and we plot these in Fig.~\ref{fig3}. The error bars in energy represent a 2$\sigma$ ($\approx 95\%$) confidence interval on the fits. The error bars in $t$ are due to the uncertainty in the measurements of the waveguide thickness. The curves overplotted with the data points in Fig.~\ref{fig3} are guides to the eye.

Figure~\ref{fig3} shows that the ratio of the coupling rate $\Omega$ to the total loss rate $\gamma_L+\gamma_G$ is maximized at an optimum waveguide thickness $t=550$ nm. For this value, $\Omega > \gamma_G$ and $\Omega \approx \gamma_L$ (within the error bar). We interpret this condition as the onset of strong coupling.  For thinner or thicker waveguides, $\Omega$ is less than at least one of the loss rates (mostly $\gamma_L$). This corresponds to the weak coupling regime, where energy dissipation is faster than energy exchange between the oscillators.  The finding that this system transitions from weak to strong coupling for a limited range of waveguide thickness is a central result of this paper.  We highlight that the system we investigate (periodic array of metallic nanoparticles coupled to a dielectric slab waveguide) has been actively studied for its ability to modify light propagation and  emission in numerous ways~\cite{Giessen03, Yannopapas09, Zentgraf09, Zhang11APL, Bernal12, Rodriguez12PRL}. While several groups have presented evidence for strong or weak coupling between LSPRs and guided modes in various configurations, this is the first time that the same plasmonic system is shown to transition between the two regimes.

Intuitively, the transition from weak to strong coupling can be explained in view of how the waveguide thickness modifies the field overlap between the TM$_0$ guided mode and the LSPR, which is localized near the base of the waveguide. In the thin waveguide limit, the guided mode is weakly confined and a significant fraction of its energy lies outside the slab. The coupling is therefore weak, because the field overlap with the nanoantennas is poor. In the thick waveguide limit, the fundamental guided mode is well confined. However, its electric field amplitude is greatest close to the center of the waveguide, far from the nanoantennas. Therefore, once again the coupling is weak because the field overlap with the nanoantennas is poor. An optimum coupling arises for an intermediate thickness, where the field overlap is greatest. In addition to this primary dependence of the coupling strength on the position of the nanoantennas, the coupling also depends on the shape of the nanoantennas. For example, we have observed (measurements not shown here) that an otherwise identical lattice of pyramidal rather than cylindrical nanoantennas displays weaker couplings with identical waveguides.  The pyramidal nanoantennas lead to significantly different emission spectra. The differences are not only attributable to  the well-known dependence of the bare LSPR energy and linewidth on the shape of the nanoantenna. We believe that also the coupling is shape-dependent because the positions of the electromagnetic hot-spots (where the field overlap with the guided mode is greatest) are shape-dependent.  While an exhaustive study of shape-dependent couplings is beyond the scope of the present paper, we hereby point to this effect for consideration in future works.

We now comment on the dependence of the loss rates on $t$. $\gamma_L$ is affected by the local density of optical states at the position of the nanoantennas. As shown by Buchler and co-workers, LSPR radiative losses are affected by a nearby dielectric interface~\cite{Buchler05}. Here, the proximity of the air-polystyrene interface to the nanoantennas (determined by $t$) leads to a  modified LSPR linewidth. This is more clearly visible in the measurements for the thinnest waveguides [see Fig.~\ref{fig2}(g)]. Besides this effect, we suspect that slightly different optical qualities (e.g. roughness) of the waveguides could also exert a small influence on our measurements.  Regarding $\gamma_G$, its non-zero value could be considered surprising based on the fact that a bare guided mode in an unstructured dielectric slab is a bound mode, which implies zero decay rate. As we explain next, $\gamma_G$ includes both radiation losses due to the structuring of the waveguide, and absorption losses due to the molecules in the waveguide.  Radiation losses are enhanced for small $t$ because the actual dye-doped polystyrene waveguide --- spatially modulated by the presence of the nanoantennas --- deviates more pronouncedly from the flat layer supporting a strictly bound mode. Our data agrees with this intuition, since Fig.~\ref{fig3} shows that $\gamma_G$ decreases as $t$ increases. At large $t$, $\gamma_G$ asymptotically approaches the absorption rate of the molecules in the waveguide ($5.3 \pm 2$ meV), which is indicated by the red dotted line in Fig.~\ref{fig3}. This absorption rate is derived from the complex refractive index of the dye-doped polystyrene layer, $\tilde{n}_p = n_p + i k_p$,  which we obtained from ellipsometric measurements. Since the ratio $n_p/k_p$  gives the number of optical cycles after which the energy density of a wave  decays, the absorption rate at frequency $\omega$ is $\gamma = k_p/n_p \omega$.  For the calculation in Fig.~\ref{fig3}, we set $\omega = \overline{\omega}_G$, where the overbar indicates an average for all measured $t$. The $\pm 2$ meV in the value quoted above represents slight variations of $\omega_G$  as a function of $t$, which change the value of $k_p$ due to the frequency dispersion of the refractive index.  It should be mentioned that a radiative contribution to $\gamma_G$ implies, by reciprocity, the possibility of direct radiative excitation of this mode.  Therefore, the assumption in our model (Eq.~\ref{COM}) that only the LSPR mode is driven directly by the harmonic force holds only approximately for small $t$, and more faithfully for large $t$.

\begin{figure*}[!]
\includegraphics[width=\linewidth]{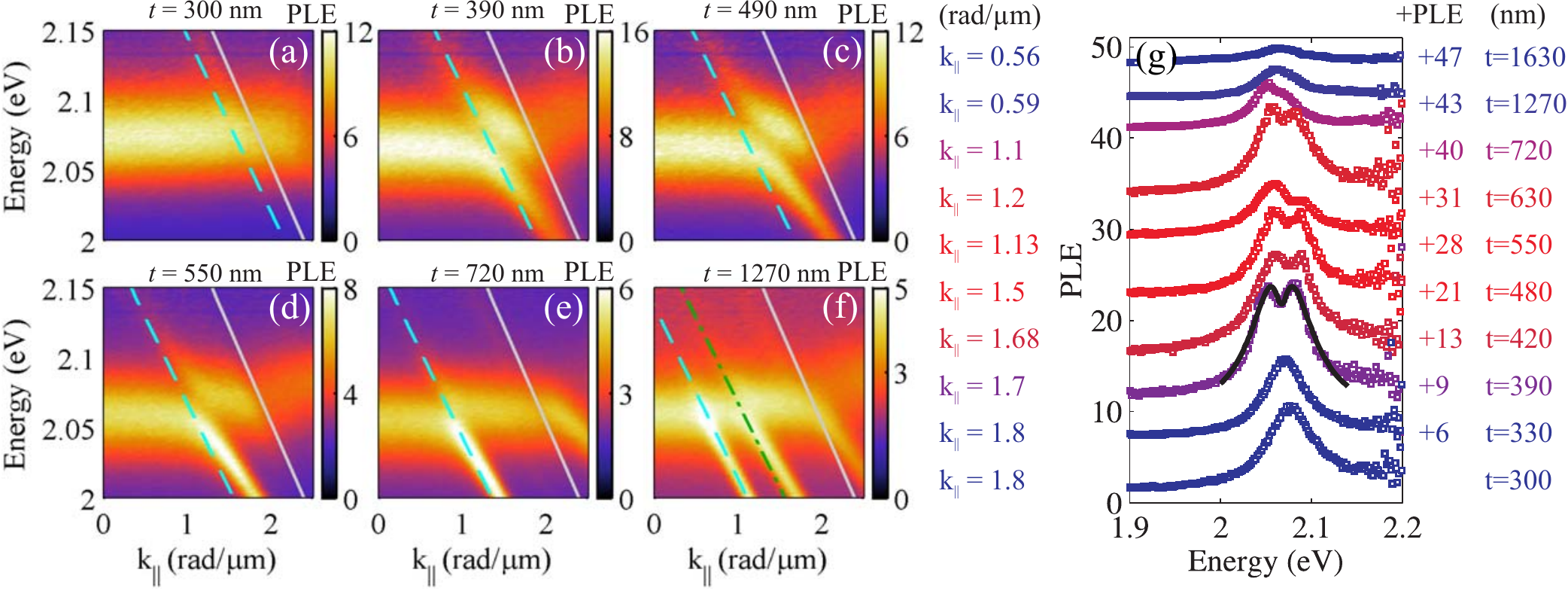}
\caption{ Photoluminescence enhancement (PLE) measurements of the structure in Fig.~\ref{fig1}, for a waveguide thickness (a) $t=300$ nm, (b) $t=390$ nm,  (c) $t=490$ nm, (d) $t=550$ nm, (e) $t=720$ nm, and (f) $t=1270$ nm. The gray solid line, identical for all plots, indicates the Rayleigh anomaly in the substrate. The cyan dashed line, shifting towards lower $k_\|$ for increasing $t$, indicates the fundamental TM$_0$ guided mode. The green dash-dotted line in (f) indicates the TM$_1$ guided mode. (g) cuts in $k_\|$ at zero detuning between the TM$_0$ guided mode and the localized surface plasmon resonance in arrays embedded in waveguides of different thickness. The value of  $k_\|$ for each cut is shown at the left of the figure. At the right we indicate the amount by which the PLE data was offset (after the ``+'' sign) and the thickness $t$ of the waveguide. The black line overplotted with the measurement for $t=390$ nm is a fit with a coupled oscillator model as described in the text.}\label{fig4}
\end{figure*}

Next we present photoluminescence measurements corresponding to the same samples discussed in  Fig.~\ref{fig2}. The samples were pumped by a 2.8 eV continuous wave laser at a fixed angle of incidence 5$^{\circ}$. The variable angle emission was collected by a fiber-coupled spectrometer rotating in the far-field, with an angular resolution of $0.2^{\circ}$. The pump irradiance ($5$ mW/mm$^2$) was far below the saturation threshold of the molecules. Figure~\ref{fig4} shows the photoluminescence enhancement (PLE) in color --- varying scales --- as a function of the emitted photon energy and $k_\|$. The PLE is defined as the ratio of the photoluminescence from the waveguide with and without the nanoantenna array.

The PLE displays an intricate dependence on $t$ that does not directly correlate with that of extinction. For $t=300$ nm [Fig.~\ref{fig4}(a)], the PLE is dominated by the LSPR  yielding a maximum 12-fold enhancement. For $t=390$ nm [Fig.~\ref{fig4}(b)], the LSPR shows weak signatures of hybridization with the TM$_0$ guided mode, while the  maximum PLE increases to roughly 16-fold. For $t=490$ nm [Fig.~\ref{fig4}(c)], the PLE from the weakly hybridized LSPR and TM$_0$ guided mode are roughly equal, reaching a maximum 12-fold enhancement. For the three thickest waveguides [Figs.~\ref{fig4}(d,e,f)] the LSPR enhancement is reduced and the PLE is dominated by the TM$_0$ guided mode. Notice that for the 4 measurements with $t>390$ nm [Figs.~\ref{fig4}(c,d,e,f)], the maximum PLE monotonically decreases.  We attribute this reduction in PLE to a  higher fraction of dye molecules that are effectively uncoupled from the nanoantenna array. For large $t$, these are the molecules near the top of the waveguide, where the nanoantenna-enhanced near-fields have significantly decayed. Note that even though the molecules are uniformly distributed throughout the waveguide, the field overlap between the optical mode and the molecules is not constant in space. In particular, as $t$ increases beyond the characteristic decay length of the nanoantenna-enhanced near-fields ($\sim$few hundred nm, depending on frequency and wave vector), the ensemble emission becomes increasingly dominated by molecules displaying a negligible field overlap with the LSPR, and little overlap with the guided mode.


We now focus on the relative strength of the PLE features and their connection to the properties of the coupled modes. We previously established, based on our analysis of the extinction spectra, that for the thinnest and thickest waveguides the system lies well into the weak coupling regime. In this case, the relevant eigenmodes are the LSPR and the TM$_0$ guided mode --- not their mixture. On either the small or large $t$ weak coupling regime, the extinction displays comparable LSPR lineshapes only marginally affected by the TM$_0$ guided mode [Figs.~\ref{fig2}(a,f)].  In contrast, the PLE differs remarkably in these two weak coupling regimes. For small $t$ the greatest contribution to the PLE comes from the LSPR [Fig.~\ref{fig4}(a)], while for large $t$ it comes from the TM$_0$ guided mode [Fig.~\ref{fig4}(e,f)].  This is due to the different decay lengths of the modes, which leads to a greater field overlap with the emitters for the LSPR at small $t$ and for the TM$_0$ guided mode at large $t$.  The extinction, on the other hand, is not affected by the field overlap of these modes with the emitters. Instead, the extinction is determined by the interference between incident and scattered fields. This leads to a spectral shift of the near-field with respect to the far-field~\cite{Bryant08, Lee09, Zuloaga11, Alonso13}, which can also explain the spectral shift of the emission enhancement with respect to the extinction in the presence of a single resonance ~\cite{Giannini10}. For coupled resonators, interference and electromagnetic retardation can lead to a more complex behavior, including a suppressed  far-field response at the same frequency, wave vector, and polarization, for which the near-field is enhanced~\cite{Hentschel10, Frimmer12, Rodriguez13OL}. While such a condition is particularly attractive for enhancing light emission with reduced losses~\cite{Rodriguez12PRL}, its relation to the weak-to-strong coupling transition has hitherto not been discussed. Here, by mapping this transition we demonstrate the different regimes in which waveguide-coupled light-emitting optical antennas can operate. On one hand, the results at small $t$ provide a design principle (optimum layer thickness) for generating angle-independent light emission enhancements. On the other hand, the results at large $t$ provide a design principle for generating directional narrowband light emission enhancements which follow the dispersion of guided modes. For intermediate $t$,  we observe that strong coupling induces a spectral window of far-field transparency accompanied by only a shallow dip in PLE at zero LSPR-guided mode detuning. Thus, in this region the near-field to far-field contrast is greatest.

We finalize the discussion of the PLE measurements by making a comparison with the measurements in Ref.~\cite{Rodriguez12PRL}, where a  nanoantenna array stands \textit{on} rather than \textit{in} a light-emitting waveguide. The greater field overlap between the optical modes enabled by the present configuration allows us to observe enhanced (but still weak) hybridization effects in PLE in the vicinity of the strong coupling regime ($390$ nm $\lesssim t \lesssim$ $630$ nm). In contrast,  no hybridization effects were observed in the PLE measurements of Ref.~\cite{Rodriguez12PRL}.  We stress the term ``weak hybridization'' because the dispersion and linewidths of the resonances are clearly modified [Fig.~\ref{fig4}(b,c,d)], but their energy splitting at zero detuning never exceeds their linewidths. This is clear qualitatively, and also quantitatively, as revealed by fitting the PLE spectrum with the same model used for the extinction spectra (Eq.~\ref{COM}). The fit to the spectrum displaying the largest splitting [black line in Fig.~\ref{fig4}(g)] yields $\Omega=29\pm 5$ meV, $\gamma_L= 75 \pm 5$ meV, and $\gamma_G= 17 \pm 7$ meV. We believe that the apparent contradiction in the values of the coupling and loss rates points to the highly interesting fact that any given system of coupled oscillators displays  distinct observables with different lineshapes depending on how the oscillators are driven. Here, for example, a time-harmonic driving of only one mode as assumed in Eq.~\ref{COM} seems inadequate to describe the PLE spectra. Recall that the PLE is generated by near-field rather than far-field excitation, and both modes can be directly excited. We also note that the maximum splitting in PLE occurs for $t=390$ nm rather than $t=550$ nm. The dependence of the apparent mode splitting on the observable quantity has been highlighted in Refs.~\cite{Savona95, Schwartz11} in view of transmission, reflection, and absorption spectra. Here, we introduce a new quantity that needs consideration in emitting systems aimed to operate in the strong coupling regime: the PLE. While an unambiguous determination of the coupling strength is in principle only possible through an eigenmode analysis, experiments always retrieve observables in a driven system. It is therefore important to understand the dependence of these observables on the key parameters of the system (e.g., $t$ in our case).  Furthermore, we point out that PLE and absorption measurements are not related through reciprocity. While Kirchoff's Law relates absorption and emission at any point in space,  an absorptance measurement of our sample largely probes the local field enhancements at the position of the nanoantennas, while PLE measurements probe  the local field enhancements at the position of the molecules. As we show next, these two quantities can differ pronouncedly depending on the coupling strength and frequency detuning of the modes supported by the structure.

In what follows, we study the transition from weak to strong coupling between the LSPR and the TM$_0$ guided mode using full wave simulations. Firstly, we confirm the features observed in experiments. Secondly, we interpret these features in terms of near-field maps.  We have used two distinct methods bench-marked against each other. These are the Fourier modal method (S$^4$) and the finite-element method (COMSOL). The Fourier modal method~\cite{Li96a} is a plane wave expansion method to calculate the transmission, reflection and diffraction of layered biperiodic discontinuous structures, i.e. stratified gratings. We use the free implementation S$^4$ of Liu and Fan~\cite{LiuFan}. We find good  convergence using  just 289 plane waves provided we use parallellogramic truncation, and employ the proper factorization rules of Li~\cite{Li96b} appropriate for high-index contrast crossed gratings. We take the same refractive index values used in the dispersion calculations ($n_{a}=1.0$, $n_{s} = 1.58$, and  $n_{g}=1.44$), and model the aluminum nanoantennas as cylinders of height 150 nm and diameter of 118 nm.  The aluminum dielectric constant we use is a polynomial parametrization of measured ellipsometry data.

To model the PLE and visualize the near-fields, we use COMSOL rather than S$^4$. The Fourier modal method is not optimized for high accuracy in fields according to a point-by-point local measure, while finite element simulations are optimal for real-space insight.  As geometry we take the same parameters as in S$^4$.  The computational domain in COMSOL  spans the unit cell in the periodicity plane, and extends several wavelengths perpendicularly into  to the substrate and superstrate. We apply Bloch-Floquet  boundary conditions at the edges of the unit cell  and periodic port conditions for the remaining domain walls.   The zero-order port on the air side is set for angled plane wave excitation. We have benchmarked the COMSOL simulations against S$^4$ by comparing the calculated extinction for the $t=300$~nm structure.  We find percent-level agreement for wave vectors below the RA in glass, i.e., in the range of the experiment. However, just beyond the RA in glass  COMSOL shows fringes in extinction, which we attribute to spurious reflections off the periodic port boundary condition that occur when a diffracted order is grazing along the port in the substrate or superstrate.  These artifacts could be reduced by extending above 7 wavelengths the computational domain in the direction perpendicular to the layers. However, this comes at the expense of an increased computational time  compared to S$^4$.  Since we find excellent correspondence for extinction at all energy and momenta below the RA in glass, we  conclude that the finite element simulation is fiducial for PLE and near-field maps in this spectral region.

\begin{figure}[!]
\includegraphics[width=\linewidth]{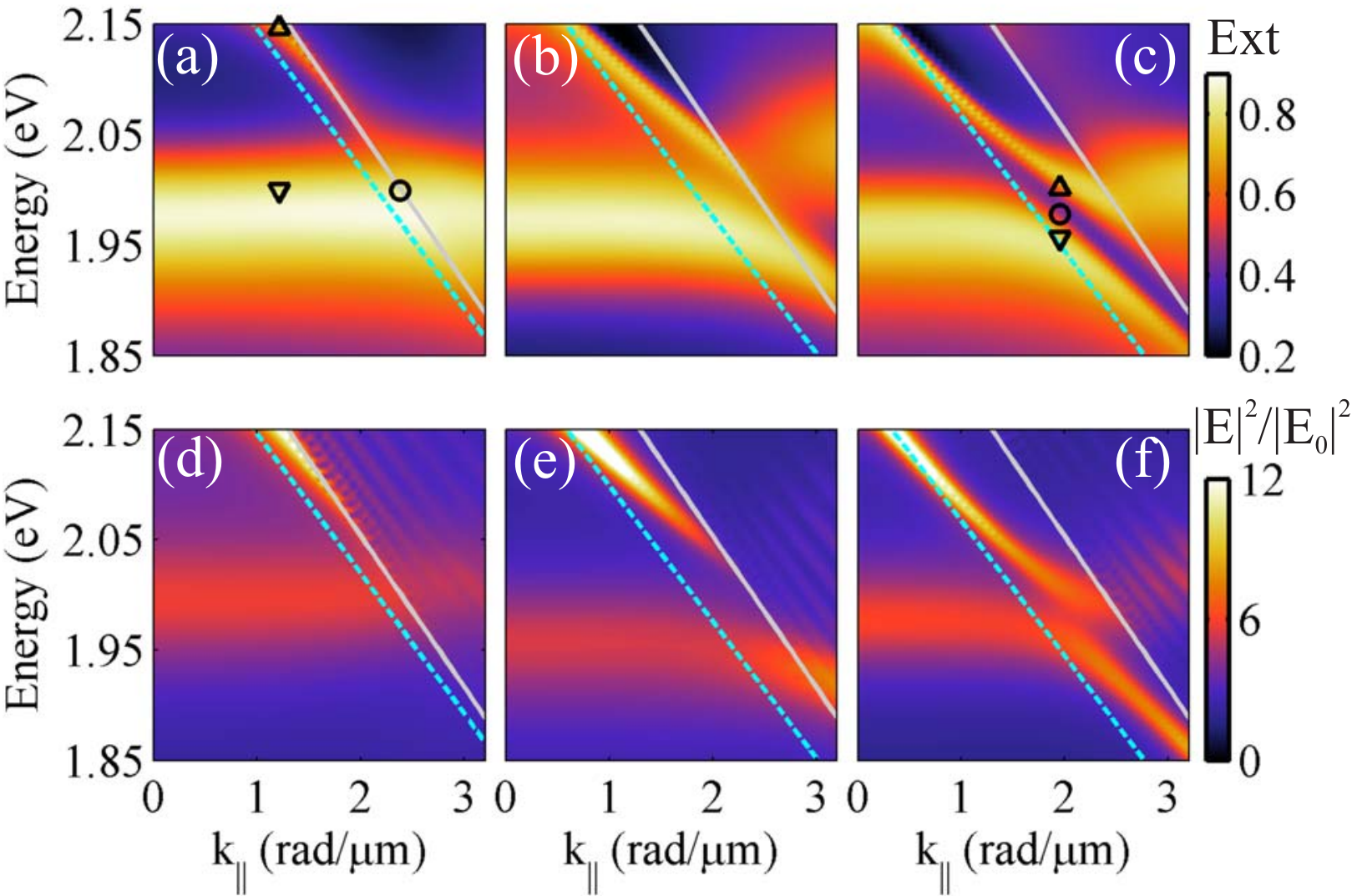}
\caption{Numerical simulations of the light extinction (a,b,c) and electric field intensity enhancement averaged over the waveguide volume (d,e,f) of  the structure shown in Fig.\ref{fig1} for different waveguide thickness $t$. For
(a,d) $t= 300$ nm, for (b,e) $t= 420$ nm, and for (c,f) $t=550$ nm.  The gray solid line, identical for all plots, indicates the Rayleigh anomaly in the substrate. The cyan dashed line, shifting towards lower $k_\|$ for increasing $t$, indicates the TM$_0$ guided mode. The open symbols in (a) and (c) indicate the energy-$k_\|$ points inspected in~\ref{fig6}.}\label{fig5}
\end{figure}

Figures~\ref{fig5}(a,b,c) show the extinction (1-T$_0$,  for TM-polarized light, incident from air) simulated with S$^4$ for three waveguide thicknesses:  (a) $t=300$ nm, (b) $t=420$ nm, and (c) $t=550$ nm.  The LSPR in the simulations is somewhat red-shifted with respect to the measurements. This is likely due to differences between the simulated and fabricated metallic structures in their dimensions or dielectric function. In addition, aluminum nanostructures present a 2-3 nanometers surface oxide (Al$_2$O$_3$) layer~\cite{Knight12}, which is not taken into account in our simulations and could be the origin of small spectral shifts.  Nevertheless, the simulations capture the essence of the measurements (Fig.~\ref{fig2}) well, both displaying a transition from weak to strong coupling as $t$ increases. Notice in Fig.~\ref{fig5}(c) that the avoided resonance crossing is centered at a larger value of $k_\|$ than expected based on the calculated guided mode dispersion, in agreement with experiments [see Fig.~\ref{fig2}(d)]. Figures~\ref{fig5}(d,e,f)  show the spectrally resolved electric field intensity enhancements for the same three waveguides, simulated with COMSOL. The enhancement is defined as $|E|^2/|E_0|^2$, with $E$ and $E_0$ the total and incident electric field, respectively, both spatially averaged over the waveguide volume (excluding the volume of the particles). $|E|^2/|E_0|^2$ is related to the PLE by reciprocity,  which states  that the local electric field enhancement in the waveguide upon far-field plane wave illumination is equivalent to the plane wave strength in the far-field due to a localized source.  Since our experiment averages all possible positions and orientations of the emitters in the entire waveguide,  we integrate the total electric field intensity enhancement over the entire volume where the emitters are located.  The resultant quantity can be regarded as the radiative part of the fractional (angle-resolved) local density of optical states. Comparing Fig.~\ref{fig5}(d) with Fig.~\ref{fig4}(a) shows that for $t=300$ nm,  the dominant contribution to the field enhancement in the waveguide comes from the LSPR. This results in a broadband PLE feature with a flat angular dispersion. For $t=550$ nm, the $|E|^2/|E_0|^2$ and PLE spectra in Figs.~\ref{fig5} (f) and Fig.~\ref{fig4}(d), respectively, display mixed features of the LSPR and guided mode with an anti-crossing between them. As in the measurements, the magnitude of the splitting at the avoided resonance crossing is smaller for PLE  than for extinction [Figs.~\ref{fig2} (d) and Fig.~\ref{fig5}(c)]. For intermediate values of $t$, the $|E|^2/|E_0|^2$ and  PLE spectra show characteristics in between these two cases. Overall, the simulated quantity $|E|^2/|E_0|^2$ qualitatively reproduces the PLE measurements. The agreement is better at lower than at higher energies because the absorption by the molecules (not taken into account in the simulations) limits the PLE at higher energies. Indeed, the imaginary component of the refractive index of the dye-doped polystyrene layer, $k_p$, is roughly a factor of four higher at 2.15 eV than at 2.06 eV (the average eigenfrequency of the TM$_0$ guided mode, $\omega_G$, as obtained from the coupled harmonic oscillator fits). Hence, we expect re-absorption of the enhanced light emission to more seriously hamper the PLE at higher energies as the waveguide thickness increases. This expectation is in agreement with our measurements in Fig.~\ref{fig4}, where the sharp feature in PLE associated with the guided mode gradually fades for energies above $\sim 2.06$ eV, and this effect becomes more pronounced as $t$ increases.

\begin{figure}[!]
\includegraphics[width=\linewidth]{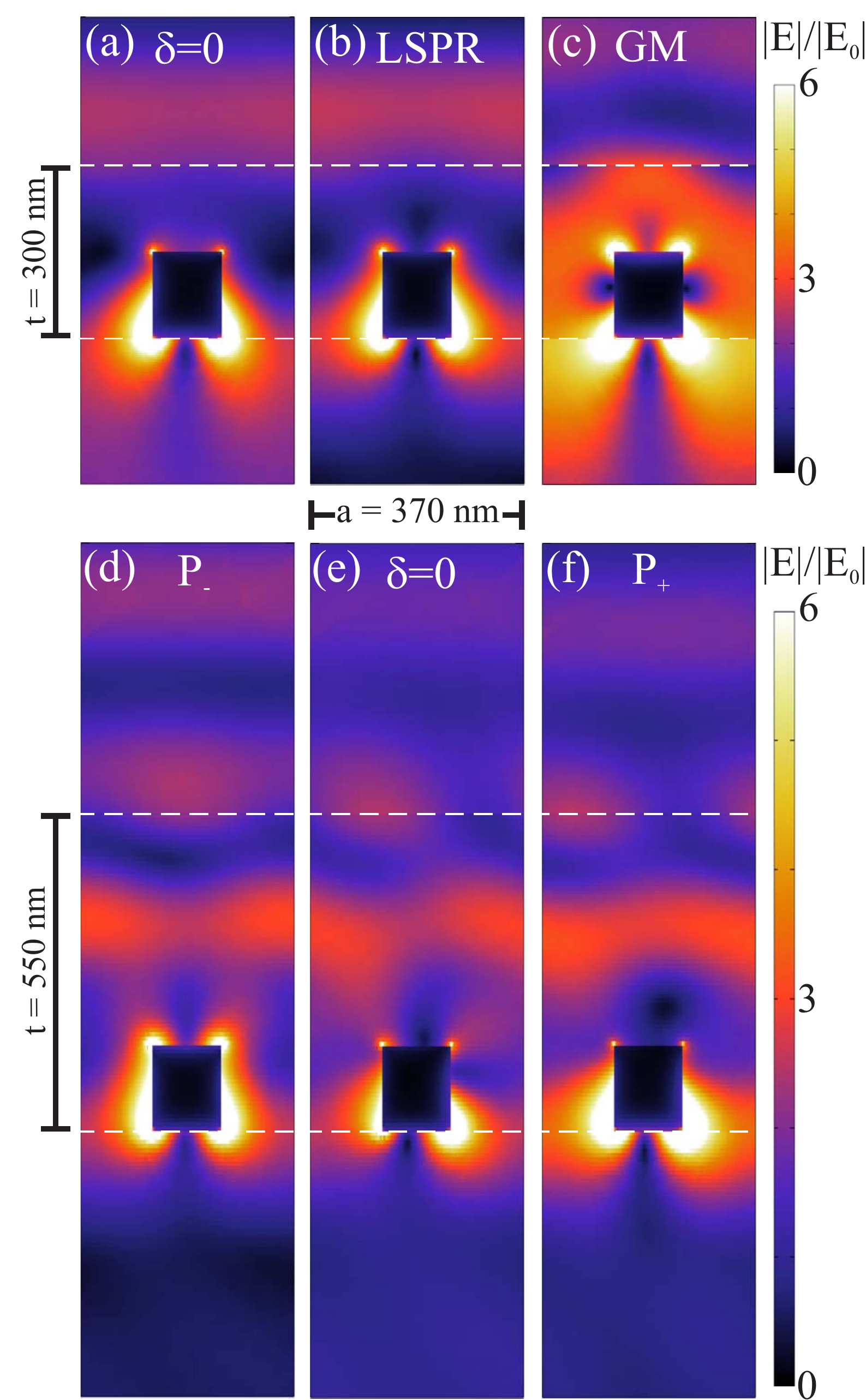}
\caption{ Electric field enhancements for the 300 nm waveguide in (a,b,c), and for the 550 nm waveguide in (d,e,f). The energy and $k_\|$ corresponding to panels (a,b,c) are indicated Fig.~\ref{fig5}(a):(a) is at the circle, (b) is at the downwards triangle, and (c) is at the upwards triangle. The energy and $k_\|$ corresponding to panels (d,e,f) are indicated in  Fig.~\ref{fig5}(c): (d) is at the downwards triangle, (e) is at the circle, and (f) is at the upwards triangle.}\label{fig6}
\end{figure}

We now inspect the near-fields of the structure at selected energies and $k_\|$  to illustrate the key differences between weak and strong coupling. In Fig.~\ref{fig6} we plot $|E|/|E_0|$ at a plane parallel to the incident electric field and intersecting the nanoantennas at their center. Figures~\ref{fig6}(a,b,c) correspond to $t=300$ nm, while Figs.~\ref{fig6}(d,e,f) correspond to $t=550$ nm. Figure~\ref{fig6}(a)  is close to zero detuning, as indicated by the open circle in Fig.~\ref{fig5}(a). Figures~\ref{fig6}(b,c) represent a large detuning, occurring at $k_\|=1.123$ rad$/\mu$m. In Fig.~\ref{fig6}(b) the photon energy is 2 eV, as indicated by the downwards triangle in Fig.~\ref{fig5}(a); this corresponds to the approximately bare LSPR. In Fig.~\ref{fig6}(c) the photon energy is 2.147 eV, as indicated by the upwards triangle in Fig.~\ref{fig5}(a); this corresponds to the approximately bare TM$_0$ guided mode. The similarity of the fields in Fig.~\ref{fig6}(a) and Fig.~\ref{fig6}(b) is due to the weak coupling, which induces a negligible modification to the LSPR even at zero detuning with the guided mode. In contrast to both Figs.~\ref{fig6}(a,b), the electric field enhancement in Fig.~\ref{fig6}(c) is stronger and more delocalized. The weaker confinement of the field to the metal explains the narrower resonance linewidth at the conditions of Fig.~\ref{fig6}(c).

Figures~\ref{fig6}(d,e,f) illustrate the near-fields for three different energies all at $k_\|=1.95$ rad$/\mu$m, which is close to zero detuning for $t=550$ nm. Strong coupling leads to two new eigenstates, which we label as P$_-$ and P$_+$ in Fig.~\ref{fig6}(d) and Fig.~\ref{fig6}(f), respectively. The energy and $k_\|$ of P$_-$ and P$_+$ are indicated by the downwards and upwards triangles in Fig.~\ref{fig5}(c), respectively. The field profiles of P$_-$ and P$_+$  are similar to each other because the strong coupling has hybridized the modes such that their individuality is lost. Here, waveguide-plasmon polaritons are a linear superposition of the bare LSPR and TM$_0$ guided modes with equal weights. If the detuning parameter is varied from  $k_\|=1.95$ rad$/\mu$m, the field solutions along the upper and lower polariton branches depart from this condition, gradually acquiring a resemblance to one or the other of the bare modes. Finally, an interesting situation occurs in Fig.~\ref{fig6}(e), the energy and $k_\|$ of which is indicated by the circle in  Fig.~\ref{fig5}(c). Here the local fields in the waveguide are still significantly enhanced but the extinction is reduced. This spectral region is  particularly attractive for light-emitting plasmonic systems, as it enables simultaneously enhanced local fields at the position of the emitters (and therefore large fluorescence enhancements) and suppressed absorption losses in the metal.

In conclusion, we have investigated the light extinction and emission angular spectra of an aluminum nanoantenna array embedded in a luminescent slab waveguide. By varying the waveguide thickness we demonstrated the transition from weak to strong coupling between localized surface plasmons in the nanoantennas and the fundamental guided mode in the slab. Our results provide a design principle for hybrid dielectric-metallic resonators aimed at improving the performance of solid-state light-emitting devices, and shed new light on the near-field to far-field contrast of optical antenna arrays. In particular, we have shown how the same nanoantenna array can provide drastically different radiation patterns in photoluminescence enhancement as the coupling strength between the aforementioned two modes is varied. From a fundamental perspective, we envisage these results to stimulate a quest for a more comprehensive description of hybrid light-matter excitations in strongly coupled systems,  as we have here shown that their observable properties (e.g. extinction and emission dispersion relations, and energy splitting of coupled modes) depend pronouncedly on the nature of the excitation source.

We thank Jorik van de Groep, Gabriel Lozano, and Erik Garnett for stimulating discussions. This work was supported by the Netherlands Foundation for
Fundamental Research on Matter (FOM) and the Netherlands Organisation for Scientific Research (NWO), and is part of an
industrial partnership program between Philips and FOM.


\end{document}